\documentstyle[11pt,newpasp,epsf]{article}

\begin{document}

\title{Round Table Summary:\\
Stellar interferometry as a tool to investigate
atmospheres and to compare observations with models}

\author{Markus Wittkowski}
\affil{European Southern Observatory, Casilla 19001, Santiago 19, Chile,
mwittkow@eso.org}






\begin{abstract}
Long-baseline interferometry at optical and near-infrared wavelengths
is an emerging technology which is quickly becoming a useful tool
to investigate stellar atmospheres and to compare observations 
with models. Stellar atmosphere models have so far mainly been 
constrained by comparisons with stellar spectra which are integrated
over the stellar disks.
Interferometric observations provide spatially and spectrally resolved 
information and can thus provide important complementary observational
information which can be compared to model predictions. 
Here, I summarize the different aspects on this topic which were discussed 
at a round table on Thursday, June 20, 2002, during IAU Symposium 210.
This summary gives an overview on discussed interferometric facilities and
techniques, concepts to study atmospheres by optical interferometry,
and particular classes of objects. We conclude that more frequent interactions
between the efforts of atmosphere modelling and interferometric observations
promise to lead to increased confidence in stellar model atmospheres
and to further advancement of the field in the next years.
\end{abstract}




%
%
%

\section{Introduction}
Long-baseline interferometry at optical and near-infrared wavelengths
is an emerging technology which is quickly becoming a useful tool
to investigate stellar atmospheres and to compare observations 
with models. A round table on this topic was held on June 20, 2002, 
during the IAU Symposium 210 ''Modelling of stellar atmospheres''. 
The goal of this round table was to discuss and stimulate interactions 
between observations obtained by optical interferometers and theoretical
modelling in the field of stellar atmospheres. Here, I summarize aspects
that were discussed at the round table, or during the course of the 
symposium. This summary is based on the individual contributions
which can be found in this volume, supplemented by some further 
literature that was cited at the symposium, without any 
claim of completeness. This round table summary is, hence, not intended
to be a complete review of the field. Optical interferometry, including 
interferometric techniques as well as scientific studies, have 
recently been reviewed by Shao \& Colavita (1992), Quirrenbach (2001), 
and Baldwin \& Haniff (2002).
OLBIN (optical long baseline interferometry news, edited by P. Lawson,
and available at olbin.jpl.nasa.gov) is an excellent source of up-to-date
information in the field of optical interferometry. Circumstellar 
envelopes are also studied by maser observations using radio 
interferometry (Humphreys, this volume). However, radio interferometry
was not part of this round table discussion. 

In general, information on stellar atmospheres using 
optical long-baseline interferometry has so far been derived by
comparing model predictions to observations and measurements of 
\begin{itemize}
\item angular sizes of stars and derived absolute radii and 
effective temperatures
\item center-to-limb intensity variations 
\item asymmetric shapes of stellar surfaces
\item horizontal surface inhomogeneities
\item circumstellar matter.
\end{itemize}
These measurements have often been temporally or spectrally
resolved, which leads to additional insights into the atmospheres'
structures and dynamics.

In the following, topics of discussion are summarized in the contexts of 
(1) interferometric facilities and techniques,
(2) concepts to study stellar atmospheres, which are
applicable to different object classes, or (3) certain classes of objects. 
There is, however, overlap between these categories.
\section{Interferometric techniques and facilities}
\paragraph{Interferometric facilities}
Results, modelling of results, or predictions using the following 
interferometric facilities were presented at the symposium.\\[2ex]
\begin{tabular}{llp{7cm}}
Name & Location & References in this volume\\ \hline\\
COAST & Cambridge, UK & Freytag \& Mizuno-Wiedner, Freytag \& Finnson,
van Belle et al. (b).\\
GI2T  & Calern, France & Dominiciano et al., Jankov et al., Stee \& Bittar.\\
IOTA & Mt. Hopkins, USA & van Belle et al. (b).\\
MARK III & Mt. Wilson, USA & Quirrenbach \& Aufdenberg.\\
NPOI & Flagstaff, USA & Wittkowski.\\
PTI  & Mt. Palomar, USA & van Belle et al. (a,b).\\
KECK & M. Kea, USA & van Belle et al. (a,b).\\
VLTI & Paranal, Chile & Dominiciano et al., Jankov et al., Stee \& Bittar, 
Wittkowski, Wittkowski et al.\\
\end{tabular}\\[2ex]
Further operating interferometric facilities include\\[2ex]
\begin{tabular}{llp{7cm}}
CHARA & Mt. Wilson, USA & \\
ISI   & Mt. Wilson, USA & \\
SUSI  & Narrabri, Australia &\\
\end{tabular}\\[2ex]
For further information on these facilities, see the
references mentioned above, Baldwin \& Haniff (2002), 
Quirrenbach (2001), or OLBIN. The above mentioned interferometric 
facilities are generally available to the stellar atmosphere community 
through acceptance of observing proposals and/or collaborations, 
with the exception of the MARK III, which is not operational anymore. 
Detailed procedures and priorities for scientific topics differ. 
For instance, ESO currently accepts scientific proposals on all 
fields from the international community to be carried out on a shared-risk 
basis with the VLTI and the near-infrared K-band commissioning instrument 
VINCI. A first call for proposals was issued for Period 70 
(October 2002 to March 2003) and a second call for Period 71 (April 2003 
to October 2003). Calls for proposals for the scientific instruments 
MIDI (mid-infrared) and AMBER (near-infrared) can be expected for 2003/2004.
In general, due to the limiting magnitudes, interferometers are mostly
used for stellar astrophysics, and hence stellar atmospheres. Additional
priorities of large facilities using 8--10m class telescopes 
(Keck, VLTI, LBT) include planet searches and active galactic nuclei.   
\paragraph{Interferometric observables}
The primary observables of an interferometer are the amplitude and phase of 
the complex visibility function, which is the Fourier transform of the
object intensity distribution (van Cittert-Zernike Theorem). 
In presence of atmospheric turbulence, usually only the squared visibility 
amplitudes and the triple products are accessible by optical and 
near-infrared interferometers. A triple product is the product of 
three complex visibilities corresponding to baselines that form a triangle.
The phase of the triple product, the closure phase, is free
of atmospheric phase noise (Jennison 1958). Thus, the measurement
of triple products is a method to obtain phase information of the
object intensity distribution, i.e. allows us to detect deviations
from point symmetry. Observational results that were
discussed at the symposium make use of squared visibility amplitudes
(Quirrenbach \& Aufdenberg, van Belle et al. (a/b), Wittkowski, 
Wittkowski et al.). The COAST images of Betelgeuse (Buscher et al. 1990),
modelled by Freytag \& Finnson and Freytag \& Mizuno-Wiedner, as well as
the NPOI limb-darkening measurements presented by Wittkowski make
in addition use of triple products.
The vast majority of interferometric studies use measurements of
relatively large (squared) visibility amplitudes before the first minimum
of the visibility function. Since this means that less than one
resolution element across the stellar disk is obtained, one parameter,
i.e. the diameter of the disk, is derived. These diameter measurements
can however be done at different position angles in order to detect
asymmetric shapes of the stellar disks (van Belle et al. (a,b)), at different
wavelengths (Quirrenbach \& Aufdenberg, Wittkowski), or at different
time periods to detect for instance stellar pulsations. 
Interferometric observations of more than one resolution element across
the stellar disks are difficult to perform, and hence rare.
The long baselines needed to obtain this resolution, i.e. to resolve the 
stellar disk, also produce very low visibility amplitudes. These 
low visibility amplitudes correspond to vanishing fringe contrasts. 
Hence, the corresponding fringes are difficult to be detected, tracked, 
and recorded. In addition, the data reduction of low visibility values 
is often a special challenge as well, since the contributions of photon 
noise and detection bias terms might become dominant. Such observations
and the successful compensation of bias terms are discussed in this volume
by Wittkowski. 

In addition to visibility amplitudes and triple products, the phase of 
the complex visibility is an interferometric observable as well, as mentioned
above. If atmospheric and instrumental noise can be controlled or compensated, 
for example by phase referencing using a reference star within the 
isoplanatic patch or by using the same source at another wavelength, 
the relative position of the object on the sky can be derived with an 
accuracy which is much higher than the diffraction limit. If the position 
of the photocenter of the object depends on the wavelength, such differential
phase measurements can be used to derive information on the spatial object
brightness distribution. Information on stellar diameters,
stellar rotation, and starspots can be derived. This technique was described 
by Beckers (1982), Chelli \& Petrov (1995), Jankov et al. (2001), and 
is discussed in this volume by  Dominiciano de Souza et al. and Jankov et al.
Jankov et al. discuss in particular the imaging potential of a tomographic
technique which combines time-resolved spectroscopy (Doppler Imaging) and
differential interferometry.

Simulations of complete atmospheres of stars, such as the
radiation hydrodynamics simulations of Betelgeuse presented by
Freytag \& Finnson and Freytag \& Mizuno Wiedner would ideally be
compared with high-resolution images of stellar surfaces derived by
interferometric observations. If modulus and phase of the complex 
visibility values are derived, the object intensity distribution can in 
principle be directly reconstructed from interferometric data using 
imaging techniques which effectively interpolate the limited coverage 
of the Fourier-plane.
A recent review describing the image fidelity using optical interferometers
can be found in Baldwin \& Haniff (2002).
In order to reconstruct an image of reasonable fidelity, it is intuitively
understandable that the number of visibility data points has to be at least as
large as the number of unknowns, i.e. the pixel intensities of the image.
Furthermore, despite the ability of imaging algorithms to effectively
interpolate the sparse aperture data, the aperture has to be filled
as uniformly as possible in order to avoid strong artifacts of the PSF
such as sidelobes. This means that many very low visibility values
at several lobes of the visibility function and at different azimuth
angles have to be measured. As a result, images will probably be limited 
to a few pixels (about 3\,$\times$\,3 -- 5\,$\times$\,5\,px.) 
in the next years. It was discussed that model fitting, for instance fitting
of parameters that describe asymmetries as a function of time, might be
a better choice for first attempts.
\paragraph{Calibration}
Interferometric observations are generally limited by the calibration of
the atmospheric and instrumental transfer function, i.e. the 
instrumental visibility response to a point source. This quantity 
is, dependent on instrument and site, observed to be dependent on 
different parameters as for instance seeing, atmospheric coherence time, or
zenith angle of the source. The transfer function is usually calibrated by
observations of an unresolved star or a star with well-known diameter.
Since, especially if long baselines are used, almost all calibrator stars
are at least slightly resolved, the uncertainty of the knowledge of the
diameter of the calibrator star becomes often a major limitation
(e.g. Percheron et al. 2002). In addition, the diameter of the 
calibrator star is often obtained at a wavelength different from 
that of the new experiment. In this case, model atmospheres need to
be used to predict the interferometric transfer function. 
This can best be done by using the Rosseland mean angular diameter. Then,
the model atmosphere is needed to derive the transfer function based on 
the Rosseland mean angular diameter, or to
transform the Rosseland mean angular diameter into a more practical
uniform disk diameter at the wavelength of the experiment.
As a result, calibration is a topic for which optical interferometry
relies on stellar atmosphere modelling, rather than providing 
observational constraints for models.
\section{Concepts to study stellar atmospheres}
The following concepts to study stellar atmospheres were discussed.
These concepts are usually applicable to different classes of 
objects.
\paragraph{Diameters, radii, effective temperatures}
The vast majority of interferometric measurements have so far directly
derived only the first stellar surface structure parameter, which is the 
angular size of the star. The combination of the angular diameter with
the bolometric flux is the most direct method to obtain the effective
temperature of the star. This is an important constraint for comparisons
of model atmospheres with stellar spectra, since it fixes one free
parameter. Another diameter that can be derived from the angular diameter
is the absolute radius of the star if its distance is known, e.g. by
using the Hipparcos parallax. Further fundamental stellar parameters
have been derived from observations of binary systems (e.g. Hummel et al.
1995). The stellar diameter is usually obtained using a uniform disk
model. The uniform disk diameter is wavelength-dependent due to the
limb-darkening effect (see, e.g. Quirrenbach et al. 1996, 
Wittkowski et al. 2001). Stellar model atmospheres are used to 
transform wavelength-dependent uniform disk diameters at the observational
wavelength into wavelength-independent limb-darkened diameters. The observed
relative variation of uniform disk diameters as a function of wavelength 
can also be compared to model predictions of this wavelength dependence,
and hence be used to test model atmospheres. Quirrenbach \& Aufdenberg
present in this volume uniform disk diameter measurements of 47 cool giants 
with the Mark III Interferometer on Mt. Wilson, California in a strong 
TiO band and in the continuum. They reproduce the diameter ratios 
by PHOENIX state of the art model atmospheres (Hauschildt et al. 1997; Baron
et al., this volume) and derive important constraints
for these models. This concept can be used for many different stars
using modern interferometers equipped with instruments that allow spectral
resolution, as for instance the VLTI with the AMBER instrument at 
near-infrared wavelengths.
Van Belle et al. (b) discuss in this volume wavelength-dependent 
diameter measurements of Mira stars. Narrowband diameters of oxygen-rich
and carbon-rich Miras and non-Mira stars obtained with the PTI are
presented; Mira diameter measurements obtained by different interferometers
are reviewed.
\paragraph{Asymmetric shapes of stellar surfaces}
Interferometric measurements can also derive asymmetric shapes of
stellar surfaces if baselines of different orientation are used.
Van Belle et al. (a) present in this volume the recent detection 
of an oblate photosphere of a fast rotator, the main sequence star
Altair (see also van Belle et al. 2001). The theoretical description 
and simulations of observations of stellar rotation are also discussed 
in this volume by Dominiciano de Souza et al. and 
Jankov et al. (see also Domiciano de Souza et al. 2002, Jankov et al. 2001).
Van Belle et al. (b) discuss in this volume that also for Mira stars
indications of departures from spherical symmetry are beginning to be
observed as increasingly rich image information is obtained by a new
generation of interferometers. Mira stars rotate slowly, and hence,
stellar rotation is an unlikely cause for these asymmetries. The nature
of the asymmetry of surfaces of AGB stars is not well understood.  
\paragraph{Intensity profiles and limb-darkening}
Optical interferometry has proven its capability to go beyond the
measurement of diameters, and to measure additional surface structure
parameters. Through the direct measurement of stellar limb-darkening,
interferometry tests the wavelength-dependent intensity profile across 
the stellar disk. Unfortunately, direct measurements of the 
limb-darkening effect are rare because of the observational difficulties
mentioned above. While diameters have so far been obtained
for several hundred stars with interferometric and lunar occultation
techniques, limb-darkening has been directly observed for only a very 
limited number of stars (Hanbury Brown et al. 1974, Di Benedetto \& Foy 1986,
Haniff et al. 1995,
Quirrenbach et al. 1996, Burns et al. 1997, Hajian et al. 1998, 
Wittkowski et al. 2001).
Wittkowski (this volume) present multi-wavelength limb-darkening 
measurements on the giant star $\gamma$\,Sge, obtained with the NPOI, 
which succeeded not only in discriminating between uniform disks 
and limb-darkened disks, but also in constraining Kurucz model atmosphere 
parameters (Wittkowski et al. 2001). First VLTI measurements of visibility 
values beyond the first minimum were shown as well.
Confronting models with both measured stellar spectra and measured intensity 
profiles at the same time is a strong test for the radiation fields predicted
by atmosphere models (see, e.g. Aufdenberg \& Hauschildt 2003).
Considering the rapid evolution of interferometry, it can be expected that
more limb-darkening parameters of a wider variety of stars can be 
measured soon with high spectral resolution.
\paragraph{Surface features}
The final goal is the reconstruction of images of stellar surfaces
with several resolution elements across the stellar disk, revealing both
the overall center-to-limb intensity variation and additional horizontal
inhomogeneities as warm or cool starspots. These inhomogeneities 
may be caused by photospheric convection, magnetic fields, or 
abundance inhomogneities. 
Small scale features on stellar surfaces have already been suggested
by interferometric observations of Capella (Hummel et al. 1994) and
$\beta$\,And (Di Benedetto \& Bonneau 1990). For the apparently 
largest supergiants $\alpha$~Orionis, $\alpha$~Scorpii and $\alpha$~Herculi, 
a few bright spots have been detected and mapped by direct imaging techniques 
including optical interferometry and HST imaging 
(Buscher et al. 1990, Wilson et al.  1992, Gilliland \& Dupree 1996, 
Burns et al. 1997, Tuthill et al.  1997, Young et al. 2000).
It is expected that starspots on relatively 
large stars can directly be detected and its parameters be constrained 
using squared visibility amplitudes and triple products obtained
with the VLTI with the AMBER instrument (see Wittkowski, this volume; 
Wittkowski et al. 2002). Jankov et al. discuss in this volume the
potential to combine the techniques of stellar differential (phase)
interferometry and Doppler imaging in order to constrain surface
features on smaller rotating stars.
Parameters describing horizontal surface inhomogeneities can be 
compared to simulations of entire stellar atmospheres. 
Freytag \& Mizuno-Wiedner and Freytag \& Finnson (this volume)
present 3D radiation hydrodynamic simulations of the envelope and
atmosphere of a red supergiant, in order to model observational data
of Betelgeuse. 
\section{Classes of objects}
A few particular classes of objects were discussed in the course of 
the symposium, for which comparisons of models with interferometric 
observational data seem to be promising.
\paragraph{Red giants and supergiants}
Red giant and supergiant stars have so far been the prime target for
interferometric observations. This is caused by their large apparent
sizes, and their brightness, especially at near-infrared wavelengths.
Surface features have been detected on the surfaces of the apparently 
largest supergiants $\alpha\,$~Orionis, $\alpha$~Scorpii and $\alpha$~Herculi
(e.g. Buscher et al. 1990, Burns et al. 1997, Young et al. 2000).
Models to describe such surface features are presented by
Freytag \& Mizuno-Wiedner and Freytag \& Finnson (this volume).
\paragraph{AGB stars and Mira stars}
The evolution of late-type stars along the asymptotic giant branch
(AGB) is characterized by high luminosities and low effective temperatures,
and accompanied by significant mass loss to the circumstellar
environment with mass-loss rates of up to $10^{-7}-10^{-4}$ M$_\odot$/year.
AGB stars with masses of the order of 1\,M$_\odot$ become unstable
to large amplitude radial pulsations and become Mira variable stars.
The processes of mass-loss and pulsation are essential for our understanding
of late stages of stellar evolution, but are still a matter of debate
(see the excellent keynote presentation by Woitke in this volume). 
Atmosphere models need to couple time-dependent dynamics, 
radiative transfer, chemistry, and dust formation (Woitke, H\"ofner et al.,
both in this volume).
Optical interferometry can provide important observational constraints
by measuring the intensity profile across the stellar disks, which
are often observed to be very different from uniform disks (e.g., Wittkowski
et al., this volume), the time/stellar phase dependent diameter
variations, and absolute radii indicating the mode of pulsation
(see, e.g. Hofmann et al. 2002). Interferometric observations of 
Mira stars have recently been performed by COAST (Young et al. 2000),
IOTA (van Belle et al. 2002, Hofmann et al. 2002), and PTI (Thompson et
al. 2002). These observations are reviewed by van Belle et al. in this
volume. Wittkowski et al. present first preliminary results of Mira
observations with the VLTI and the VINCI commissioning instrument.
\paragraph{Small cool stars}
Ludwig (this volume) present model atmospheres for mid M-type
main-, as well as pre-main-sequence objects. These model atmospheres
are characterized by low effective temperatures and high
surface gravities. Because of the rather small angular sizes, these
objects are difficult for interferometric observations. However, first
measurements of diameters of M dwarfs succeeded already 
(Lane et al. 2001, Segransan et al. 2003). These observations
provide an empirical mass-radius relation, and can thus provide
observational constraints for atmosphere models of this parameter
space (low masses, low effective temperatures, high surface gravities). 
\paragraph{Winds from Be stars}
Classical Be stars are rapidly rotating hot giant or main-sequence stars.
They show asymmetric circumstellar winds in the light of emission lines.
Stee \& Bittar (2001, and this volume) present theoretical models of 
Be star winds, and predict constraints by near-infrared and visible 
interferometric observations. Optical interferometry has already been
successfully used to study Be star winds (Thom et al. 1986, Quirrenbach
et al. 1993, Stee et al. 1995). Upcoming facilities with high
spatial and high spectral resolution as for instance VLTI with the AMBER
instrument are expected to provide stronger constraints on Be star winds.
\section{Concluding remarks}
Stellar atmosphere models are still mostly constrained by observations
of stellar spectra only, which are integrated over the stellar disks. 
This round table discussion has shown that optical interferometry has 
already proven its ability to provide important additional 
observational constraints by spatially resolved observations of stellar
disks and circumstellar matter. The comparison of model
atmospheres with both stellar spectra and interferometric observations
is a strong test and leads as well to increased confidence in these models. 
Observational results for different object classes and for different 
parameter spaces of stellar model atmospheres have been presented
in the course of this symposium.
Optical interferometry is currently rapidly evolving, and simulations show
that stronger constraints for more stars can be expected soon.
Most progress in this field can be made by increasing interactions
between the efforts of theoretical atmosphere modelling and interferometric
observations.   
\acknowledgments
I am grateful to the speakers of this round table discussion,
Bernd Freytag, Robert Kurucz, Andreas Quirrenbach, Slobodan Jankov, 
Philippe Stee, Gerard van Belle, for their contributions.
I thank the SOC for allocating time for a round table on stellar
interferometry, and for the invitation as a round table organizer.
%
%
%
%

%

\end{document}